\documentclass[epjCONF,columns]{svjour} 
\usepackage{graphics}
\usepackage[varg]{txfonts} 
\usepackage[latin1]{inputenc}
\session-title{Hadron Collider Physics Symposium 2011}
\def\pt{\ensuremath{p_{\mathrm{T}}}}

\newcommand{\mct}{\ensuremath{m_\mathrm{CT}}}
\newcommand{\Mmes}{\ensuremath{M_{\mathrm{mes}}}}
\def\chinopm{\ensuremath{\mathchoice%
      {\displaystyle\raise.4ex\hbox{$\displaystyle\tilde\chi^\pm$}}%
         {\textstyle\raise.4ex\hbox{$\textstyle\tilde\chi^\pm$}}%
       {\scriptstyle\raise.3ex\hbox{$\scriptstyle\tilde\chi^\pm$}}%
 {\scriptscriptstyle\raise.3ex\hbox{$\scriptscriptstyle\tilde\chi^\pm$}}}}
\def\chinoonepm{\ensuremath{\mathchoice%
      {\displaystyle\raise.4ex\hbox{$\displaystyle\tilde\chi^\pm_1$}}%
         {\textstyle\raise.4ex\hbox{$\textstyle\tilde\chi^\pm_1$}}%
       {\scriptstyle\raise.3ex\hbox{$\scriptstyle\tilde\chi^\pm_1$}}%
 {\scriptscriptstyle\raise.3ex\hbox{$\scriptscriptstyle\tilde\chi^\pm_1$}}}}
\def\ninoone{\ensuremath{\mathchoice%
      {\displaystyle\raise.4ex\hbox{$\displaystyle\tilde\chi^0_1$}}%
         {\textstyle\raise.4ex\hbox{$\textstyle\tilde\chi^0_1$}}%
       {\scriptstyle\raise.3ex\hbox{$\scriptstyle\tilde\chi^0_1$}}%
 {\scriptscriptstyle\raise.3ex\hbox{$\scriptscriptstyle\tilde\chi^0_1$}}}}
\def\ninotwo{\ensuremath{\mathchoice%
      {\displaystyle\raise.4ex\hbox{$\displaystyle\tilde\chi^0_2$}}%
         {\textstyle\raise.4ex\hbox{$\textstyle\tilde\chi^0_2$}}%
       {\scriptstyle\raise.3ex\hbox{$\scriptstyle\tilde\chi^0_2$}}%
 {\scriptscriptstyle\raise.3ex\hbox{$\scriptscriptstyle\tilde\chi^0_2$}}}}
\def\squark{\ensuremath{\tilde{q}}}
\def\gluino{\ensuremath{\tilde{g}}}
\def\stop{\ensuremath{\tilde{t}}}

\def\sbottomL{\ensuremath{\tilde{b}_{\mathrm{L}}}}
\begin{document}
\title{SUSY Searches at ATLAS}
\author{Paul de Jong\inst{1}\fnmsep\thanks{\email{paul.de.jong@nikhef.nl}}, on behalf of the ATLAS Collaboration}
\institute{Nikhef, P.O.Box 41882, NL-1009 DB Amsterdam}
\abstract{
Recent results of searches for supersymmetry by the ATLAS collaboration
in up to $2$ fb$^{-1}$ of $\sqrt{s} = 7$ TeV $pp$ collisions at the
LHC are reported.
} 
\maketitle
\section{Introduction}
\label{intro}
Due to the high centre-of-mass energy of $7$ TeV, the LHC has
discovery potential for new heavy particles beyond the Tevatron
limits
even with little luminosity. This holds in particular for particles
with colour charge, such as squarks and gluinos in 
supersymmetry (SUSY)~\cite{susy}.
However, due to the excellent luminosity performance of the LHC in
2011, sensitivity also exists for electroweak production of
charginos and neutralinos, the supersymmetric partners of the
electroweak gauge bosons and the Higgs boson. In this document,
a number of results of ATLAS searches for supersymmetry with up
to $2$ fb$^{-1}$ of LHC $pp$ data at $\sqrt{s} = 7$ TeV are
summarized. Since none of the analyses have observed any excess
above the Standard Model expectation, limits on SUSY parameters or
masses of SUSY particles are set.
It is, however, important to consider carefully the assumptions
made in each of the limits, and the true constraints that they
impose on supersymmetry.

\section{Searches with jets and missing momentum}
\label{sec:1}
Assuming conservation of R-parity, the lightest supersymmetric particle
(LSP) is stable and weakly interacting, and will typically escape 
detection. If the primary produced particles are squarks or
gluinos (and assuming a negligible lifetime of these particles),
this will lead to final states with energetic jets
and significant missing transverse momentum.

ATLAS carries out analyses with a lepton veto~\cite{0lepton}, 
requiring one isolated lepton~\cite{1lepton}, or requiring two
or more leptons~\cite{2lepton}. In addition, a dedicated search
is performed for events with high jet multiplicity~\cite{multijet}.
Data samples corresponding to luminosities between $1.0$ and $1.3$
fb$^{-1}$ are used.
Events are triggered either on the presence of a jet plus large
missing momentum, or on the presence of at least one high-$\pt$
lepton. Backgrounds to the searches arise from Standard Model
processes such as vector boson production plus jets ($W$ + jets,
$Z$ + jets), top quark pair production and single top production,
QCD multijet production, and diboson production. Backgrounds are
estimated in a semi-data-driven way, using control regions in
combination with a transfer factor obtained from simulation.

\begin{figure}
\resizebox{0.95\columnwidth}{!}{%
  \includegraphics{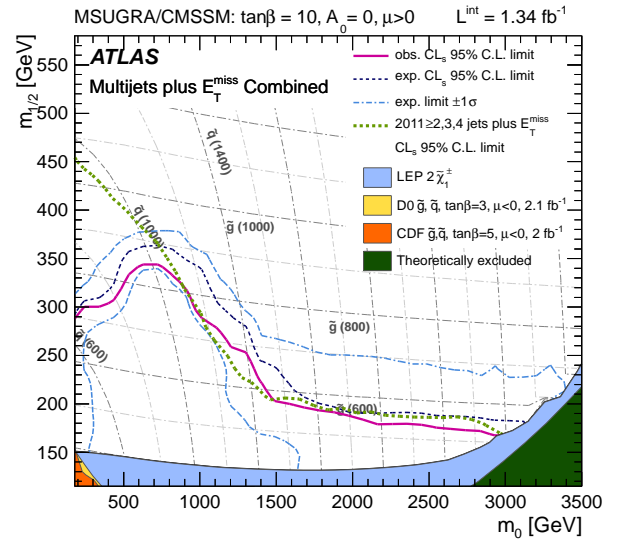} }
\caption{Exclusion contours in the MSUGRA/CMSSM $m_0$-$m_{1/2}$
plane for $A_0 = 0$, $\tan \beta = 10$ and $\mu > 0$, arising
from the analysis with $\geq2$, $\geq3$ or $\geq4$ jets plus
missing momentum, and the multijets plus missing momentum analysis.}
\label{fig:msugra}       
\end{figure}

The results are interpreted in the MSUGRA/CMSSM model, and in
particular as limits in the plane spanned by the common scalar
mass parameter at the GUT scale $m_0$ and the common gaugino 
mass parameter at the
GUT scale $m_{1/2}$, for values of the common trilinear coupling
parameter $A_0 = 0$, Higgs mixing parameter $\mu > 0$, and ratio
of the vacuum expectation values of the two Higgs doublets
$\tan \beta = 10$. Figure~\ref{fig:msugra} shows the results 
for the analyses with $\geq2$, $\geq3$ or $\geq4$ jets plus
missing momentum, and the multijets plus missing momentum analysis.
For a choice of parameters leading to equal squark and gluino
masses, squark and gluino masses below approximately $1$ TeV are excluded. 
The 1-lepton and 2-lepton
results are less constraining in MSUGRA/CMSSM for this choice of
parameters, but these analyses are complementary, and therefore
no less important.

\begin{figure}
\resizebox{0.95\columnwidth}{!}{%
  \includegraphics{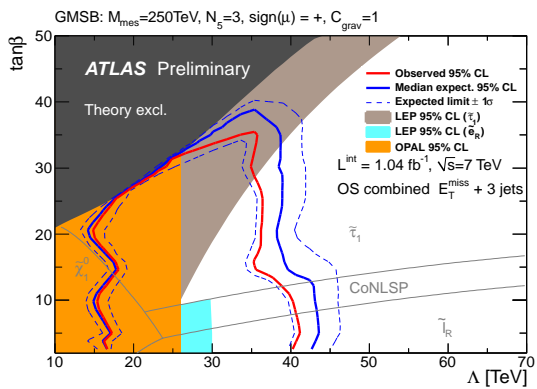} }
\caption{Exclusion contours in the GMSB $\Lambda$-$\tan \beta$
plane, for $\Mmes = 250$ TeV, $N_5 = 3$ and $\mu > 0$,
resulting from the opposite-charge lepton analysis.}
\label{fig:gmsb}       
\end{figure}

The search with two isolated opposite-charge leptons is also interpreted 
in the framework of minimal gauge mediated supersymmetry breaking
(GMSB), as shown in Figure~\ref{fig:gmsb}~\cite{gmsb}.
Assuming a messenger mass scale $\Mmes$ of $250$ TeV, 3 generations
of messengers ($N_5 = 3$) and $\mu > 0$, limits are set on the effective
SUSY breaking scale $\Lambda$ and on $\tan \beta$. These limits
significantly improve on the LEP results.

\section{Simplified model interpretation}
\label{sec:2}

ATLAS has found it useful to not only interpret the results
in constrained models, but also in terms of simplified models
assuming specific production and decay 
modes~\cite{simplemodels}. In such simplified
models, the constraints implied by models like MSUGRA/CMSSM
or GMSB are relaxed, leaving more freedom for variation of
particle masses and decay modes.
Interpretations in simplified models thus show better the
limitations of the analyses as a function of the relevant 
kinematic variables, and aid in drawing conclusions from
the results.

Inclusive search results with jets and missing momentum are
interpreted using simplified models with either pair
production of squarks or of gluinos, or production of
squark-gluino pairs.
Direct squark decays ($\squark \to q \ninoone$) or direct gluino
decays ($\gluino \to q \bar{q} \ninoone$) are dominant if all other
particle masses have multi-TeV values, so that those do not play
a role. Additional complexity may be built in, for example by
allowing one-step decays to intermediate charginos, $\chinopm$,
or heavier neutralinos, $\ninotwo$.

\begin{figure}
\resizebox{0.95\columnwidth}{!}{%
  \includegraphics{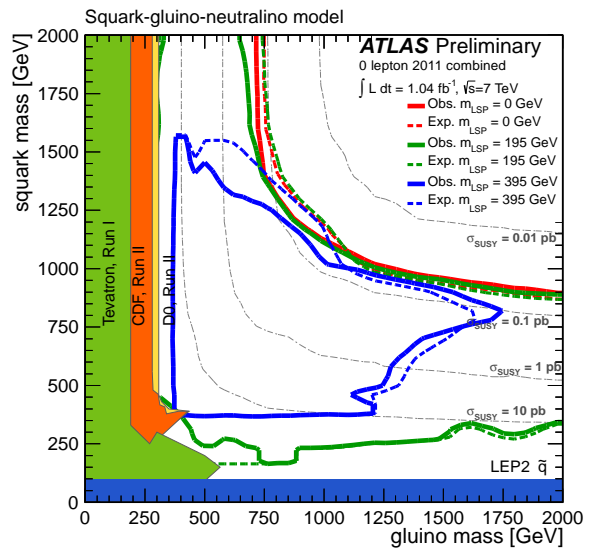} }
\caption{Exclusion contours in the squark-gluino mass plane, for
three values of the LSP mass, using the simplified model
described in the text.}
\label{fig:msqmgl}       
\end{figure}

\begin{figure}
\resizebox{0.95\columnwidth}{!}{%
  \includegraphics{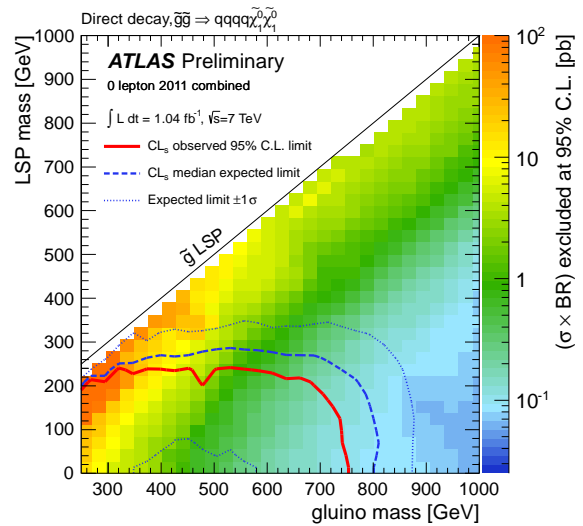} }
\caption{Cross section limits and exclusion contours in the 
gluino-neutralino mass plane, for direct gluino decays,
$\gluino \to q \bar{q} \ninoone$,
as obtained by the no-lepton analysis.
All squark masses are assumed to be multi-TeV, so that only
gluino pair production takes place, and the direct decay is
assumed to occur with 100\% branching fraction.}
\label{fig:gllsp}       
\end{figure}

\begin{figure}
\resizebox{0.95\columnwidth}{!}{%
  \includegraphics{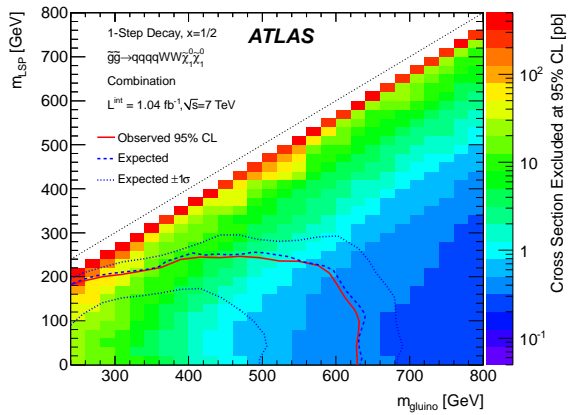} }
\caption{Cross section limits and exclusion contours in the 
gluino-neutralino mass plane, for one-step gluino decays, 
$\gluino \to q \bar{q}' \chinopm$, $\chinopm \to W^{(*)} \ninoone$,
as obtained by the one-lepton analysis. Only gluino pair 
production is considered, the one-step gluino decay is assumed 
to occur with 100\% branching fraction, and the chargino
mass is characterized by $x = 1/2$ (see text).}
\label{fig:onestep}       
\end{figure}

Figure~\ref{fig:msqmgl} shows the ATLAS results interpreted
in terms of limits on (first and second generation) squark
and gluino masses, for three values of the LSP ($\ninoone$)
mass, and assuming that all other SUSY
particles are very massive~\cite{simplified}.
Further interpretations are done in terms of limits on gluino 
mass vs LSP mass assuming high squark masses,
as shown for example in Figure~\ref{fig:gllsp} for direct
decays, or in terms of limits on squark mass vs LSP mass 
assuming high gluino masses~\cite{1lepton,simplified}.
Figure~\ref{fig:onestep} shows an example of limits in the
gluino-LSP mass plane obtained from one-step gluino decays,
$\gluino \to q \bar{q}' \chinopm$, $\chinopm \to W^{(*)} \ninoone$,
by the one-lepton analysis. The chargino mass in such decays
is a free parameter, characterized by $x = (m_{\chinopm} -
m_{\ninoone})/(m_{\gluino} - m_{\ninoone})$, and
Figure~\ref{fig:onestep} shows $x = 1/2$ as an example.

The results of the inclusive jets plus missing momentum searches,
interpreted in these simplified models,
indicate that masses of first and second generation squarks
and of gluinos must be above approximately $750$ GeV. 
An important caveat in this
interpretation is the fact that this is only true for neutralino
LSP masses below approximately $250$ GeV (as in MSUGRA/CMSSM for
values of $m_{1/2}$ below ${\cal{O}}(600)$ GeV). For higher LSP 
masses, the squark
and gluino mass limits are significantly less restraining.
It will be a challenge for further analyses to extend the
sensitivity of inclusive squark and gluino searches to the
case of heavy neutralinos. If the LSP is heavy, events are
characterized by less energetic jets and less missing transverse
momentum. This will be more difficult to trigger on, and lead
to higher Standard Model backgrounds in the analysis.

\section{SUSY and naturalness}
\label{sec:3}

Important motivations for electroweak-scale supersymmetry are the
facts that SUSY might provide a natural solution to the hierarchy
problem by preventing ``unnatural'' fine-tuning of the Higgs sector,
and that the lightest stable SUSY particle is an excellent dark
matter candidate. It is instructive to consider what such
a motivation really requires from SUSY: a relatively light top
quark partner (the stop, $\stop$) (and an associated 
sbottom-left quark, $\sbottomL$), a gluino not much heavier than 
about $1.5$ TeV to keep the stop light (the stop receives
radiative corrections from loops like $\stop \to \gluino t \to \stop$), 
and electroweak gauginos below the TeV scale~\cite{barbieri}. There are
no strong constraints on first and second generation squarks and
sleptons; in fact heavy squarks and sleptons make it easier for
SUSY to satisfy the strong constraints from flavour physics.
Motivated by these considerations, ATLAS explicitly searches for
third generation squarks and for electroweak gauginos.

\section{Stop and sbottom searches}
\label{sec:4}

ATLAS has carried out a number of searches for supersymmetry
with $b$-tagged jets, which are sensitive to sbottom and stop
quarks production, either direct, or in gluino decays.
Jets are tagged as originating from $b$-quarks by an algorithm
that exploits both track impact parameter and secondary vertex
information.

\begin{figure}
\resizebox{0.95\columnwidth}{!}{%
  \includegraphics{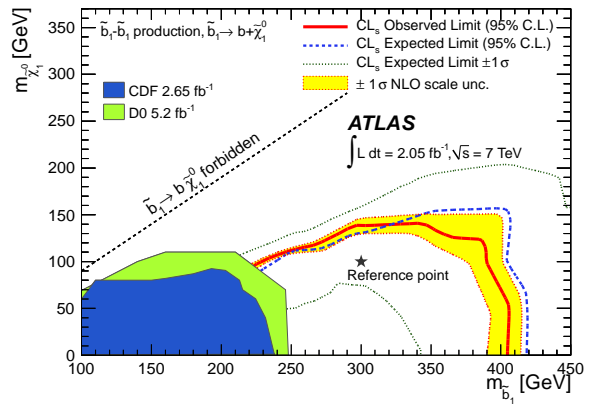} }
\caption{Exclusion contour in the sbottom-neutralino mass
plane resulting from the analysis searching for sbottom quark
pair production, assuming sbottom to bottom plus neutralino decay.}
\label{fig:sbottom}       
\end{figure}

Direct sbottom pair production is searched for in a data
sample corresponding to $2$ fb$^{-1}$ by requiring
two $b$-tagged jets with $\pt > 130, 50$ GeV and significant missing 
transverse momentum of more than $130$ GeV~\cite{sbottom}. 
The final discriminant is the boost-corrected contransverse 
mass $\mct$~\cite{mct}, and signal regions with $\mct > 100,
150, 200$ GeV are considered. No excesses are observed above
the expected backgrounds of top, $W$+heavy flavour and $Z$+heavy
flavour production. Figure~\ref{fig:sbottom} shows
the resulting limits in the sbottom-neutralino mass plane, assuming
sbottom quark pair production and sbottom quark decay into a
bottom quark plus a neutralino (LSP) with a 100\% branching fraction.
Under these assumptions, sbottom masses up to $390$ GeV are excluded
for neutralino masses below $60$ GeV.

\begin{figure}
\resizebox{0.95\columnwidth}{!}{%
  \includegraphics{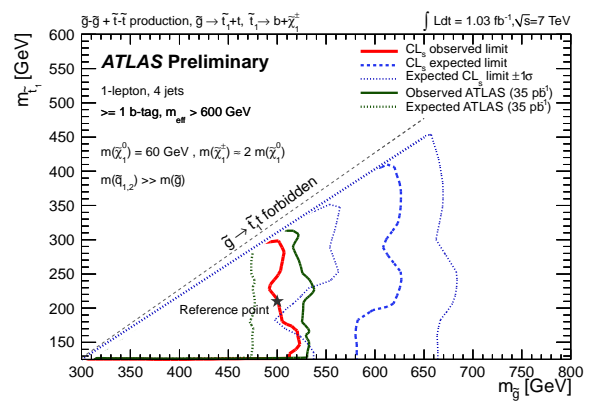} }
\caption{Exclusion contour in the gluino-stop mass
plane resulting from the analysis searching for stop
quark production in gluino decays. The assumptions
made to derive the plot are listed in the plot.}
\label{fig:glstop}       
\end{figure}

ATLAS has searched for stop quark production in
gluino decays~\cite{stop} using an analysis requiring
at least four high-$\pt$ jets of which at least one should
be $b$-tagged, one isolated lepton, and significant missing
transverse momentum. After applying the selection criteria,
$74$ events are observed in $1.0$ fb$^{-1}$ of data, where
$55 \pm 14$ background events are expected from a data-driven
estimation procedure, or $52 \pm 28$ from Monte Carlo simulations.
Since there is no significant excess, limits are set in the
gluino-stop mass plane, assuming the gluino to decay as
$\gluino \to \stop t$, and the stop quark to decay as
$\stop \to b \chinoonepm$, as shown in
Figure~\ref{fig:glstop}.

In addition, ATLAS has searched for sbottom production in
gluino decays, setting limits in the gluino-sbottom mass
plane and in the gluino-neutralino mass plane~\cite{gluinosbottom}.

Further searches for direct stop quark pair production are in progress.
These searches are challenging due to the similarity with the
top quark pair production final state for stop masses similar to
the top mass, and due to the low cross section for high stop
masses. ATLAS has searched for signs of new phenomena in
top quark pair events with large missing transverse
momentum~\cite{toppartner}; such an analysis is sensitive to
pair production of massive partners of the top quark, decaying to
a top quark and a long-lived undetected neutral particle.
No excess above background was observed, and limits on the
cross section for pair production of top quark partners are set.
These limits constrain fermionic exotic fourth generation quarks,
but not yet scalar partners of the top quark, such as the stop
quark.

\section{Electroweak gaugino searches}
\label{sec:5}

Searches for charginos and neutralinos are carried out via
analyses of final states involving photons plus missing momentum,
or multileptons plus missing momentum.

\begin{figure}
\resizebox{0.95\columnwidth}{!}{%
  \includegraphics{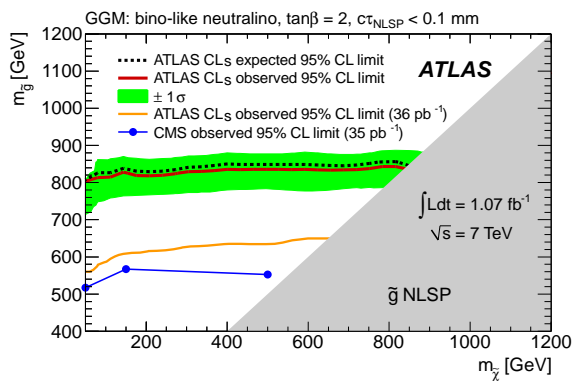} }
\caption{Exclusion contour in the gluino-neutralino mass
plane in the general gauge mediation (GGM) model,
assuming a bino-like neutralino, resulting from the
diphoton plus missing momentum analysis.}
\label{fig:ggm}       
\end{figure}

In gauge mediation models, neutralinos decay to gravitinos plus
one or more standard model particles, depending on the neutralino
composition. For bino-like neutralinos, the final state consists
of a pair of high-$\pt$ photons plus missing transverse momentum.
ATLAS has searched for an excess in such final states using
$1.1$ fb$^{-1}$ of data~\cite{diphoton}. The selection requires two 
photons, identified with ``tight'' criteria, with $\pt > 25$ GeV, 
and significant missing transverse momentum.
The results are interpreted in the general gauge mediation 
model (GGM), in terms of limits in the gluino-neutralino
mass plane, and assuming the neutralino to be the NLSP.
The results are shown in Figure~\ref{fig:ggm}.
The assumption is made that photons are produced promptly,
i.e. $c \tau$ of the NLSP is assumed to be less than $0.1$ mm.
In this model, a gluino mass below $805$ GeV is excluded for
bino masses above $50$ GeV.

The diphoton plus missing transverse momentum analysis is also
interpreted in the minimal gauge mediation model (GMSB), for
the SPS8 parameters $\Mmes = 2 \Lambda$, $N_5 = 1$,
$\tan \beta = 15$ and $\mu > 0$. The ATLAS results imply
a lower limit on $\Lambda$ for the SPS8 parameters of
$145$ TeV at 95\% CL.

\begin{figure}
\resizebox{0.99\columnwidth}{!}{%
  \includegraphics{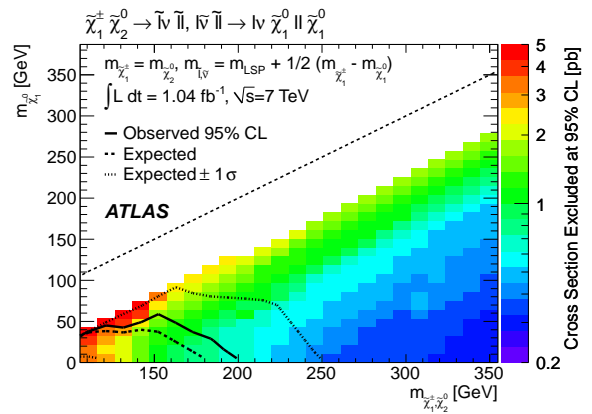} }
\caption{Cross section limits and exclusion contours in the 
chargino-neutralino mass
plane, resulting from the same-charge dilepton analysis.
The interpretation is done in a simplified model, details
are given in the plot.}
\label{fig:ewlimits}       
\end{figure}

Multilepton analyses~\cite{2lepton,multilepton} are sensitive 
to production of charginos and/or neutralinos other than the LSP, decaying
leptonically to the LSP. These analyses comprise the golden search
modes at the Tevatron, but are also rapidly gaining relevance at
the LHC. ATLAS searches for excesses in final states with three or more 
leptons on the 2011 data are in progress. ATLAS has published results
of various analyses searching for dilepton events plus missing momentum,
in $1.0$ fb$^{-1}$ of data~\cite{2lepton}. Three signal regions are
defined for opposite-charge leptons, and two signal regions are defined
for same-charge leptons, with varying selection criteria on jets and
on the missing transverse momentum. For all signal regions, the observed
event count agrees with the expected background. The analysis
selecting same-charge leptons plus large missing momentum
is sensitive to electroweak gaugino production, and results for this
analysis are shown in Figure~\ref{fig:ewlimits}. The interpretation is
done in a simplified model assuming chargino ($\chinoonepm$) plus a
heavier neutralino ($\ninotwo$) production, and decay to leptons 
and LSPs through intermediate sleptons. Under the assumption of
equal mass of $\chinoonepm$ and $\ninotwo$, limits are set
in the $\chinoonepm - \ninoone$ mass plane.

\section{Special final states}
\label{sec:6}

The number of different final states sensitive to SUSY
production is very large. SUSY particles
may be long-lived, when their decay is suppressed kinematically
(split SUSY, R-hadrons, anomaly-mediated SUSY breaking, certain
parts of phase space of gauge-mediated SUSY breaking)
or by very small couplings (e.g. R-parity violation). 
ATLAS has carried out searches for
stable massive particles~\cite{longlived}, for stopped
gluinos~\cite{stopped}, for kinked or 
disappearing tracks~\cite{kinked} and for secondary vertices
of decaying massive particles~\cite{2ndvertex}. Furthermore,
there is a dedicated search for third generation sneutrinos
decaying to an electron-muon pair in R-parity violation
scenarios~\cite{emuresonance}. It is also noteworthy that
ATLAS has searched for a scalar partner of the gluon~\cite{sgluon}.

\begin{figure}[h]
\resizebox{0.95\columnwidth}{!}{%
  \includegraphics{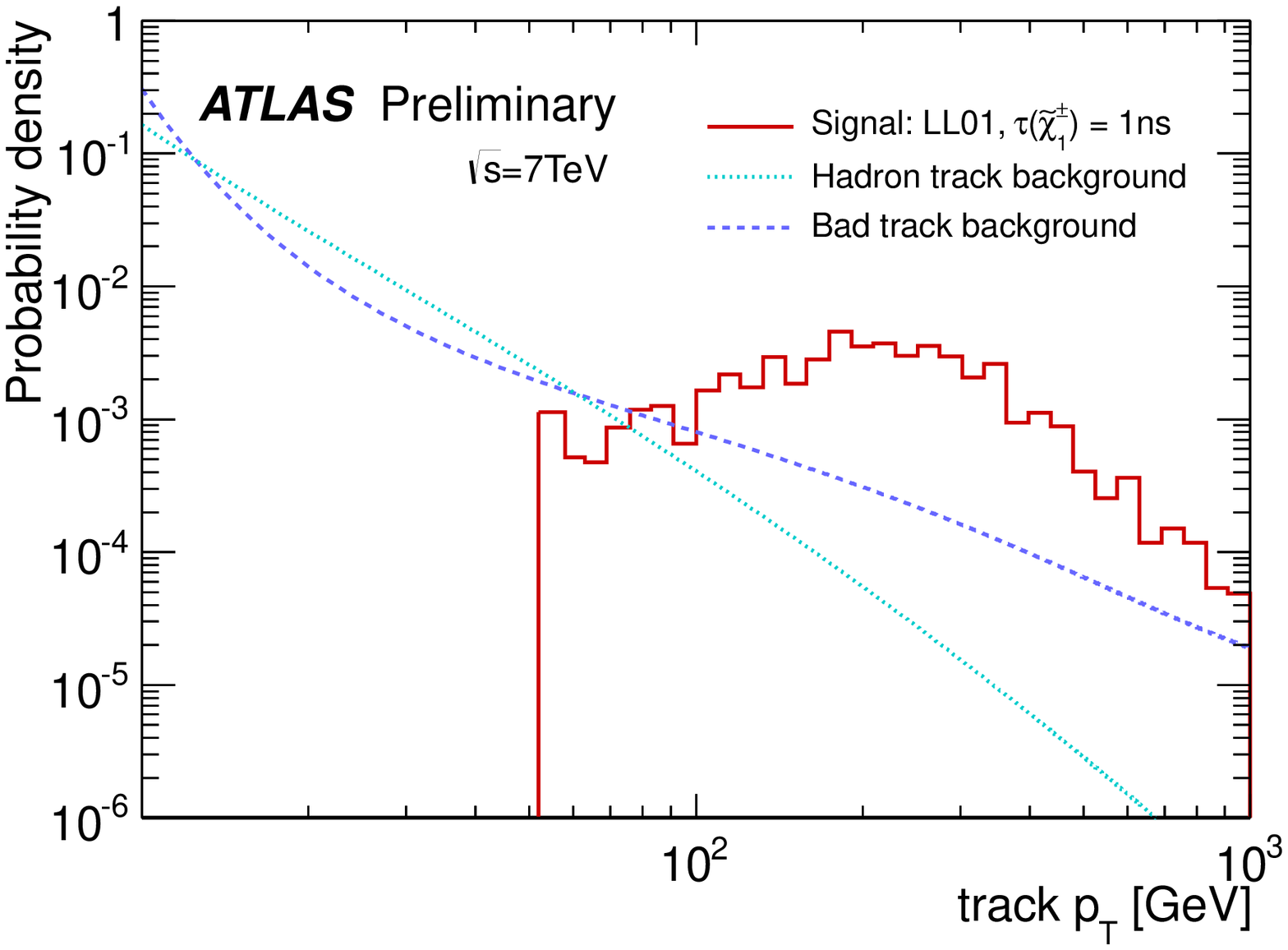} }
\resizebox{0.95\columnwidth}{!}{%
  \includegraphics{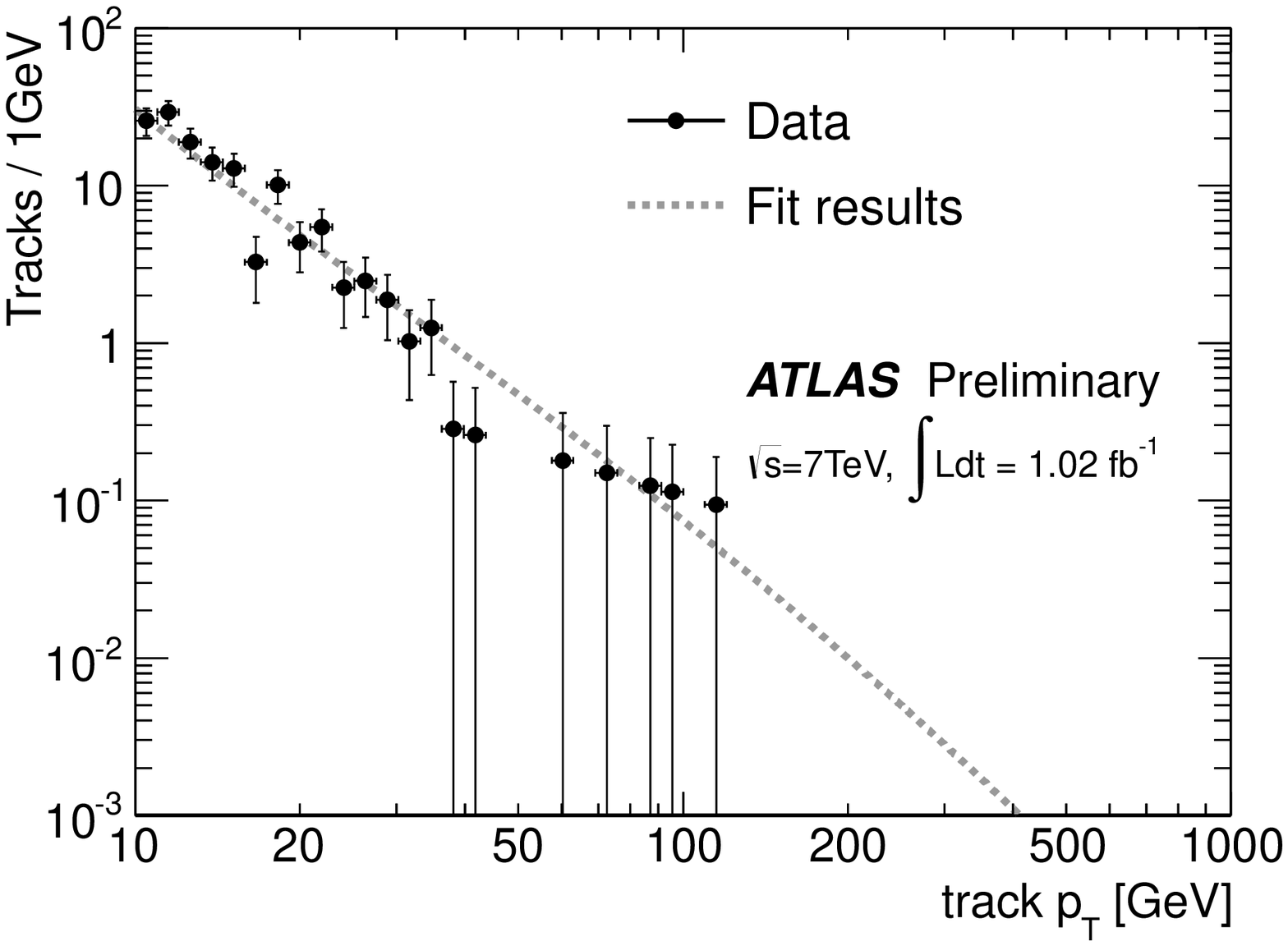} }
\caption{Top: Probability density functions for signal (AMSB
model) and expected background distributions of track $\pt$ for
tracks satisfying the kinked-track selection.
Bottom: Distribution of track $\pt$ for $185$ tracks in
data satisfying the kinked-track selection, and the result
of the pdf fit. The fit is consistent with the background-only
hypothesis.}
\label{fig:kinked}       
\end{figure}

Kinked or disappearing tracks are a possible signature of
high-$\pt$ massive particles decaying in the detector volume
to an almost degenerate daughter particle, such as
$\chinoonepm \to \ninoone \pi^{\pm}$ in anomaly-mediated SUSY
breaking (AMSB) models, where $\chinoonepm$ and $\ninoone$ are almost
degenerate, and the resulting pion track has low $\pt$ and
is easily missed in the reconstruction. 
ATLAS has searched for such signatures in
$1.0$ fb$^{-1}$ of data~\cite{kinked}, demanding a track $\pt$ of at least
$10$ GeV, good reconstruction quality in the silicon tracking
detectors and in the inner layers of the transition radiation
tracker (TRT), but no, or only few hits in the outer layer
of the TRT. Backgrounds arise from tracks interacting with the
TRT material (dominant), or from misreconstructed low-$\pt$
tracks. Figure~\ref{fig:kinked} (top) shows probability density
functions (pdfs) in $\pt$ for signal and background tracks; 
Figure~\ref{fig:kinked} (bottom) shows the $\pt$ distribution of the
$185$ tracks in data satisfying the selection criteria, and
the pdf fit to the data. The data is consistent with the
background expectation, and upper limits on the signal are set.

\begin{figure}[h]
\resizebox{0.95\columnwidth}{!}{%
  \includegraphics{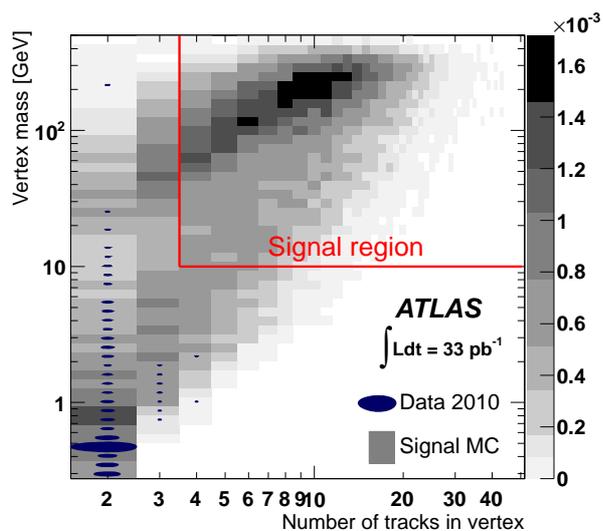} }
\caption{Vertex mass and number of tracks in the secondary
vertex, for vertices selected by the analysis searching
for high mass secondary vertices.}
\label{fig:2ndvertex}       
\end{figure}

ATLAS has also searched for high-mass secondary vertices,
consistent with the decay of massive particles, in $33$ pb$^{-1}$
of data collected in 2010. The analysis is designed in particular
for the decay $\tilde{\chi}^0 \to \tilde{\mu} \mu$ and the
R-parity violating decay $\tilde{\mu} \to q \bar{q}'$ through a
non-zero $\lambda'_{2ij}$ coupling~\cite{2ndvertex}. 
Backgrounds arise from interactions in the inner detector
material, and the fiducial volume of this analysis excludes regions
with such detector material. A signal region is defined
requiring a vertex mass of $10$ GeV or more, with at least
four tracks in the vertex, as shown in Figure~\ref{fig:2ndvertex}.
The data is consistent with the background hypothesis.

\section{Conclusion and Outlook}
\label{sec:7}

\begin{figure*}
\resizebox{0.9\textwidth}{!}{%
  \includegraphics{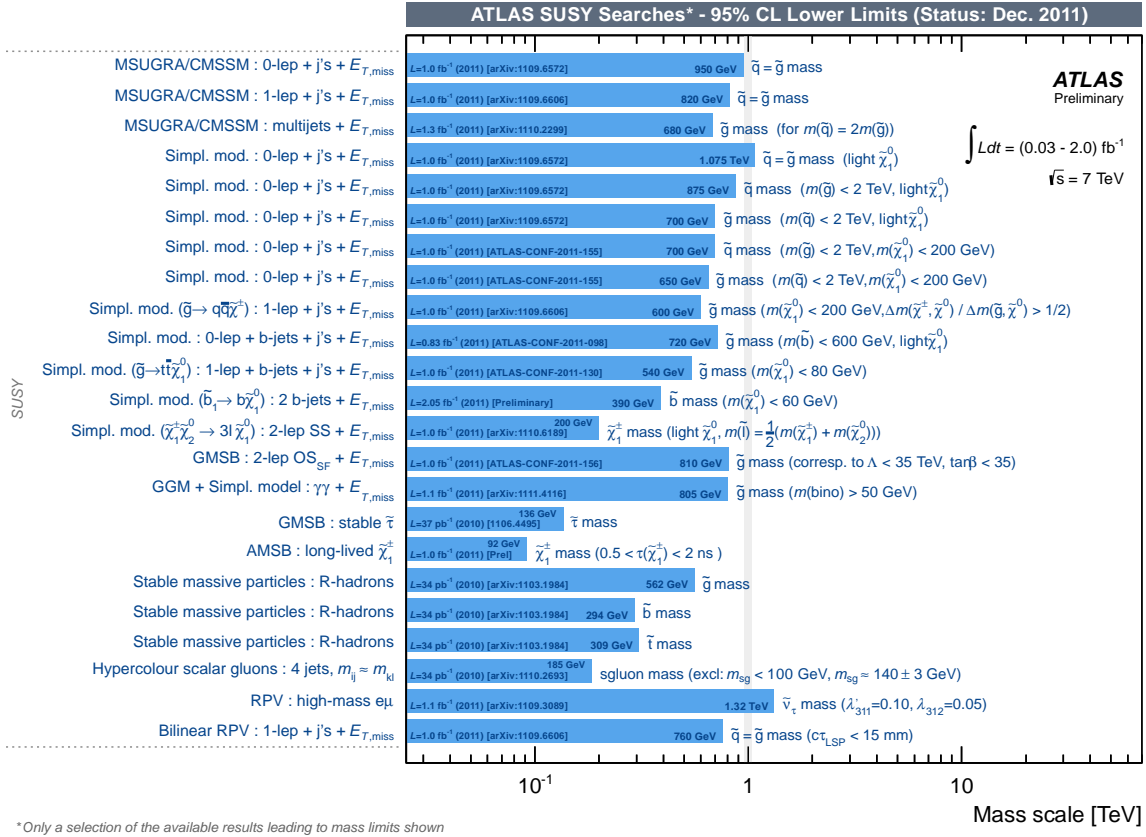} }
\caption{Summary of limits set on SUSY particle masses
by ATLAS, resulting from analyses of up to
$2$ fb$^{-1}$ of $pp$ collision data at $\sqrt{s} = 7$ TeV.}
\label{fig:summary}       
\end{figure*}

The results of ATLAS supersymmetry searches are summarized in
Figure~\ref{fig:summary}. 

Although no signs of SUSY have been
found so far, it is important to realize that actual tests of
``natural'' SUSY are only just beginning~\cite{polesello}. 
In this respect,
the LHC run of 2012, with an expected luminosity of more
than $10$ fb$^{-1}$, possibly at $\sqrt{s} = 8$ TeV, will be
very important. However, experimentally there will be considerable
challenges in triggering, and in dealing with high pile-up
conditions. In the longer term, increasing the LHC beam energy
to $>6$ TeV will again enable the crossing of kinematical barriers
and open the way for multi-TeV SUSY searches.

%

%
%


\begin{thebibliography}{}
\bibitem{susy}
See the reviews of S. P. Martin, hep-ph/9709356 and H. P. Nilles,
Phys. Rept. {\bf 110} (1984) 1, and references therein.
\bibitem{0lepton}
ATLAS Collaboration, arXiv:1109.6572 (2011)
\bibitem{1lepton}
ATLAS Collaboration, arXiv:1109.6606 (2011)
\bibitem{2lepton}
ATLAS Collaboration, arXiv:1110.6189 (2011)
\bibitem{multijet}
ATLAS Collaboration, JHEP {\bf 11} (2011) 99
\bibitem{gmsb}
ATLAS Collaboration, ATLAS-CONF-2011-156 (2011), http://cdsweb.cern.ch/record/1398247/
\bibitem{simplemodels}
D. Alves et al., arXiv:1105.2838 (2011)
\bibitem{simplified}
ATLAS Collaboration, ATLAS-CONF-2011-155 (2011), http://cdsweb.cern.ch/record/1398201/
\bibitem{barbieri}
R. Barbieri, these (HCP 2011) proceedings
\bibitem{sbottom}
ATLAS Collaboration, arXiv:1112.3832 (2011)
\bibitem{mct}
G. Polesello and D. Tovey, JHEP {\bf 03} (2010) 030
\bibitem{stop}
ATLAS Collaboration, ATLAS-CONF-2011-130 (2011), http://cdsweb.cern.ch/record/1383833/
\bibitem{gluinosbottom}
ATLAS Collaboration, ATLAS-CONF-2011-098 (2011), http://cdsweb.cern.ch/record/1369212/
\bibitem{toppartner}
ATLAS Collaboration, arXiv:1109.4725 (2011)
\bibitem{diphoton}
ATLAS Collaboration, arXiv:1111.4116 (2011)
\bibitem{multilepton}
ATLAS Collaboration, ATLAS-CONF-2011-039 (2011), http://cdsweb.cern.ch/record/1338568/
\bibitem{longlived}
ATLAS Collaboration, Phys. Lett. {\bf B 703} (2011) 428,
Phys. Lett. {\bf B 701} (2011) 1
\bibitem{stopped}
ATLAS Collaboration, (in preparation)
\bibitem{kinked}
ATLAS Collaboration, (in preparation)
\bibitem{2ndvertex}
ATLAS Collaboration, arXiv:1109.2242 (2011)
\bibitem{emuresonance}
ATLAS Collaboration, Eur. Phys. J. {\bf C 71} (2011) 1809
\bibitem{sgluon}
ATLAS Collaboration, Eur. Phys. J. {\bf C 71} (2011) 1828
\bibitem{polesello}
G. Polesello, these (HCP 2011) proceedings
\end{thebibliography}
\end{document}